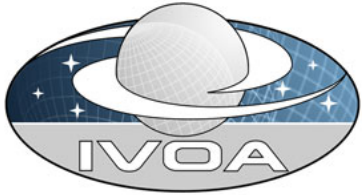

# IVOA Web Services Basic Profile Version 1.0

## IVOA WG Recommendation 2010 December 16

**This version:**
 http://www.ivoa.net/Documents/WSBasicProfile/20101216
**Latest version:**
 http://www.ivoa.net/Documents/WSBasicProfile/
**Previous versions:**
 PR: http://www.ivoa.net/Documents/WSBasicProfile/20102010
 PR: http://www.ivoa.net/Documents/WSBasicProfile/20100226
 WD: http://www.ivoa.net/Documents/WSBasicProfile/20080916

**Working Group:**
 http://www.ivoa.net/twiki/bin/view/IVOA/IvoaGridAndWebServices
**Author(s):**
 Andre Schaaff
 Matthew Graham (Editor)

## Abstract

This document describes rules to take into account when implementing SOAP-based web services. It explains also how to check conformance to these rules. It can be read as a "Guideline to VO Web Service Interoperability" or a "How to provide interoperable VO web services".

## Status of this Document

Recommendation

This document has been produced by the IVOA Grid and Web Services Working Group. It has been reviewed by IVOA Members and other interested parties, and has been endorsed by the IVOA Executive Committee as an IVOA Recommendation as of 16 December 2010. It is a stable document and may be used as reference material or cited as a normative reference from another document. IVOA's role in making the Recommendation is to draw attention to the specification and to promote its widespread deployment. This enhances the functionality and interoperability inside the Astronomical Community.

A list of current IVOA Recommendations and other technical documents can be found at http://www.ivoa.net/Documents/.

## Acknowledgments

This document has been developed with support from the National Science Foundation's Information Technology Research Program under Cooperative Agreement AST0122449 with The Johns Hopkins University, from the UK Particle Physics and Astronomy Research Council (PPARC), from the European Commission's (EC) Sixth Framework Programme via the Optical Infrared Coordination Network (OPTICON), and from EC's Seventh Framework Programme via its eInfrastructure Science Repositories initiative.

## Conformance-related definitions

The words "MUST", "SHALL", "SHOULD", "MAY", "RECOMMENDED", and "OPTIONAL" (in upper or lower case) used in this document are to be interpreted as described in IETF standard, RFC 2119 [6].

The **Virtual Observatory (VO)** is a general term for a collection of federated resources that can be used to conduct astronomical research, education, and outreach. The **International Virtual Observatory Alliance (IVOA)** is a global collaboration of separately funded projects to develop standards and infrastructure that enable VO applications. An International Virtual Observatory (IVO) application is an application that takes advantage of IVOA standards and infrastructure to provide some VO service.

## Contents

## 1. Introduction

The use of SOAP-based web services is common in the VO, e.g., footprint and spectrum services [7], SkyNodes and Open SkyQuery [8], registry interfaces [9] and CDS access [10]. New service providers need to know how to use the existing industry specifications for SOAP-based services in an IVOA context. This guideline should be an "interoperability guarantee" for the future. Our goal is not to create this guideline from scratch but to base it on existing work, specifically that of the Web Services Interoperability (WS-I) organization [11]. This document describes which parts of the existing WS-I profiles we want to use, which parts we want to replace with our own work and which parts are missing and where we want to add.

### 1.1 The Role in the IVOA Architecture

The IVOA Architecture [12] provides a high-level view of how IVOA standards work together to connect users and applications with providers of data and services, as depicted in the diagram in Fig. 1.

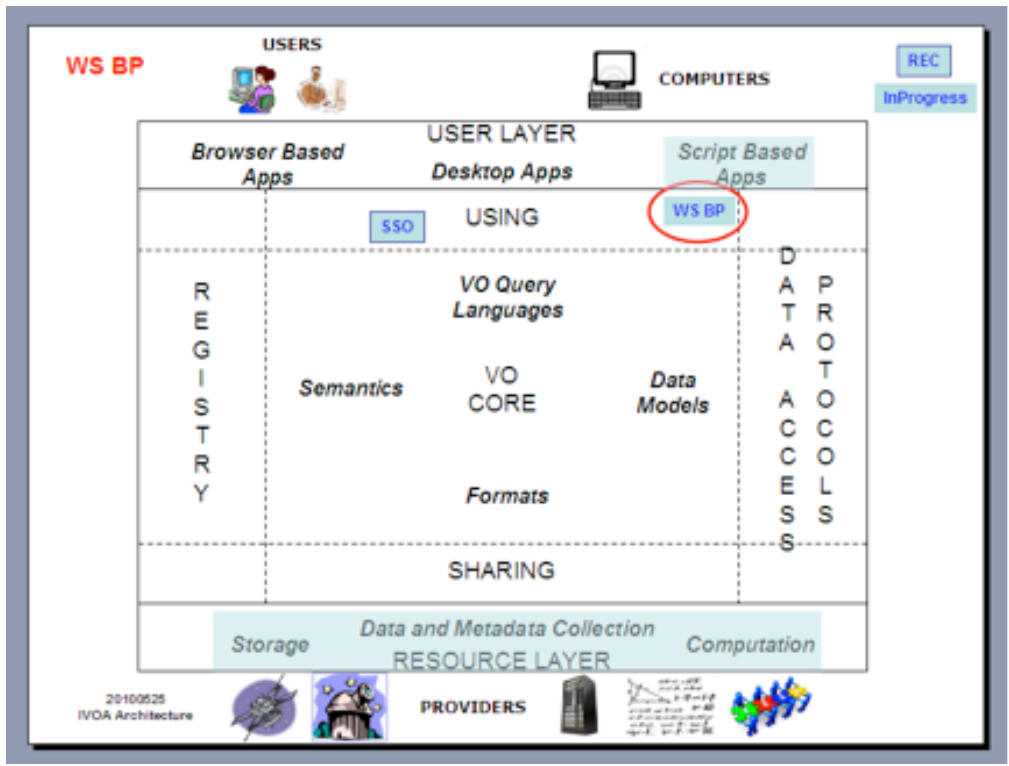

**Figure 1. WS Basic Profile in the IVOA Architecture.** This provides industry-backed
guidelines on constructing SOAP-based web services so that they are interoperable.

In this architecture, users employ a variety of tools (from the User Layer) to discover and
access archives and services of interest (represented in the Resource Layer). WS Basic
Profile ensures that all SOAP-based interfaces that are part of this usage are
interoperable.

## 1.2 Purpose

The Web Services Interoperability [11] organization is an open industry effort chartered to
promote web service interoperability across platforms, applications, and programming
languages. Its role is not to develop new specifications (like the W3C, for example) but to
interpret the existing ones and to explain how to make them work together in the best
way.

The WS-I Basic Profile is a set of non-proprietary web service specifications (SOAP,
WSDL, UDDI, XML, XML Schema, ...). It provides clarifications because using a
specification is very well but using it correctly and in the same way than others is better
for good interoperability. Specifications are also often ambiguous.

WS-I Basic Profile is supported by major world companies and working groups. For
example:

- On Microsoft web pages [13]: "Microsoft applauds the ratification of the Basic
  Profile 1.0..."

- On Apache Axis web pages [14]: "For Axis 1.2, we are focusing on our
  document/literal support to better address the WS-I Basic Profile 1.0..."

The purpose of this specification is to define an IVOA Web Services Basic Profile based
on the WS-I Basic Profile that takes into account VO standards and practices relating to
SOAP-based web services.

## 1.3 SOAP-based web services

SOAP (Simple Object Access Protocol) [15] is an XML-based messaging framework
used for information exchange between peers (client and service) in a distributed
environment. It defines the message structure but not the message content.

A SOAP message consists of a SOAP envelope that encloses the SOAP header and the
SOAP body. The header is optional and contains information about the contents of the
body, such as routing, transactional, security, contextual or user profile information. The
body contains the actual data (message payload) to be consumed and processed by the receiver of the message.

## 2. WS-I Profiles

This section discusses the various WS-I profiles and how the VO is addressing them.

### 2.1 WS-I Basic Profile

The WS-I Basic Profile 1.1 [16] describes:

- Messaging: exchange of web service protocol elements
- Description: enumeration of the messages associated with a web service, with implementation details
- Discovery: metadata which gives information about the web service
- Security: mechanism which provides integrity, confidentiality, and authentication

As far as discovery is concerned, the IVOA has decided not to adopt UDDI and so this part may be considered as being replaced by the work carried out by the IVOA Registry WG [9].

Security aspects are also covered by work carried out by the IVOA Grid and Web Services WG and this is described in Section 5.

**2.1.1 WS-I Basic Profile format**

In each part of the profile (HTTP, SOAP binding, etc.), recommendations are explained using the following syntax:

**Rxxxx** statement text

For example:

**R0001** *An instance of a web service MUST be defined by a WSDL service description*

**R1140** *A message SHOULD be sent using HTTP/1.1*

**R1141** *A message MUST be sent using either HTTP/1.1 or HTTP/1.0*

Before each rule or set of rules, the document explains the context and justifies the rule creation. The rules are not all at the same level; the compliance to one rule can be mandatory and the compliance to another can be optional.

### 2.2 WS-I Simple SOAP Binding Profile

WS-I Basic Profile 1.0 and its errata are equivalent to WS-I Basic Profile 1.1 and the WS-I Simple SOAP Binding Profile [17]. The Simple SOAP Binding Profile 1.0 is a "subset" of the Basic Profile 1.0 requirements related to the serialization of the SOAP envelope and its representation in the message.

VO web services shall be checked that they follow this profile when they do not use attachments.

### 2.3 WS-I Attachment Profile

The WS-I Attachment Profile [18] adds support for sending interoperable attachments with SOAP messages. It defines a MIME (Multipurpose Internet Mail Extensions) multipart structure for packaging attachments with SOAP messages. However, it is not mandatory to associate this profile with the WS-I Basic Profile 1.1.

WS-I has chosen the most common solution but it is too restrictive and not mandatory to associate this profile with the WS-I Basic Profile 1.1.

Although MIME is the most common solution, it is generally regarded as too restrictive. Alternate industry solutions, such as DIME (Direct Internet Message Encapsulation) [19] and MTOM (Message Transmission Optimization Mechanism, inside attachments) [20], have been defined and there is varying support for these from software libraries, such as Axis.

In light of this, this document does not define an attachment profile and VO service providers may select whichever solution best suits their requirements.

### 2.4 WS-I Basic Security Profile

The basic security profile for VO services is defined by the IVOA Single-Sign-On Profile [21]. This replaces the VS-I Basic Security Profile [22].

## 3. WS-I Testing Tools

The VO has not implemented any conformance checking software of its own but relies instead on tools provided by industry for this purpose. In this section we briefly explain which tools we have used to check conformance with the WS-I standards.

### 3.1 Available tools

The WS-I has developed conformance testing tools [23] to check services against the rules of the WS-I Basic Profile. The first provided tool is a Monitor and Analyzer package. These tools are based on configuration files which allow the user to enable/disable rules to test (assertion files).

### 3.2 Tool validation

Within the VO, the conformance testing tools have been tried with services implemented using both Tomcat/Axis and .NET infrastructures. This has been done against both the WS-I Basic Profile 1.0 and the WS-I Basic Profile 1.1 and WS-I Simple SOAP Binding 1.0 specifications.

### 3.3 Use restrictions

Due to a very restrictive license, these tools and their related libraries cannot be integrated into other tools and applications. They are, however, available to download and standalone usage with both Java and C# implementations.

## 4. The IVOA Web Service Basic Profile

This section formally defines the IVOA Web Service Basic Profile employing the same syntax as the WS-I Basic Profile for consistency.

### 4.1 WS-I conformance

VO SOAP-based web services must be compliant with WS-I specifications as follows:

*R0001* An IVOA SOAP-based web service **MUST** be compliant with the WS-I Basic Profile 1.1 [16]

*R0002* An IVOA SOAP-based seb service **MUST** be compliant with the WS-I Simple SOAP Binding 1.0 [17]

These rules may be validated through the WS-I Testing tools [23].

### 4.2 IVOA Support Interfaces conformance

VO SOAP-based web services must be compliant with the IVOA Support Interfaces (VOSI) specification [24] as follows:

*R0100* The "getCapabilities" interface **SHALL** return a valid document describing the metadata of this service.

*R0110* A VO service **SHALL** implement the "getAvailability" interface.

*R0111* The "getAvailability" interface **SHALL** return an XML document as defined in the VOSI specification [24]

*R0120* A VO service returning tabular data **SHALL** implement an interface for retrieving table metadata.

*R0121* The interface for returning table metadata **SHALL** return an XML document as defined in the VOSI specification [24]

Note that VOSI does not explicitly define a SOAP binding so that SOAP-based web services must employ the defined REST binding.

These rules may be validated through the VOSI validator.

#### 4.2.1 Assertion definition for VOSI conformance

The tool developed to test the conformance with the IVOA Support Interfaces is based on a similar assertion definition than the WS-I tools. It can be easily evolved in the future to test conformance to other IVOA standards.

```
<?xml version="1.0" encoding="UTF-8"?>
```

```
<vowsbasicprofile version="1.0">
  <description>
    This document contains the test assertions for the VOSI component
of the IVOA WS Basic Profile
  </description>

  <testAssertion id="R0100" type="required" enabled="true">
    <context>Support Interfaces</context>
    <assertionDescription> capabilities interface testing
    </assertionDescription>
    <failureMessage>the "capabilities" interface is missing</failureMessage>
    <failureDetailDescription>an IVOA Web Service should provide a "capabilities" interface </failureDetailDescription>
    <testToDo>
      <InterfaceChecking>capabilities</InterfaceChecking>
    </testToDo>
  </testAssertion>

  <testAssertion id="R0110" type="required" enabled="true">
    <context>Support Interfaces</context>
    <assertionDescription>capabilities interface return testing
    </assertionDescription>
    <failureMessage>capabilities do not return an XML file conform to the VOSICapabilities schema</failureMessage>
    <failureDetailDescription>the "capabilities" interface shall return an XML document conform to the VOSICapabilities schema</failureDe
    <testToDo>
      <OutputChecking type="XSD" interface="capabilities">link to VOSICapabilities schema </OutputChecking>
    </testToDo>
  </testAssertion>

  <testAssertion id="R0200" type="required" enabled="true">
    <context>Support Interfaces</context>
    <assertionDescription> availability interface testing
    </assertionDescription>
    <failureMessage>the "availability" interface is missing</failureMessage>
    <failureDetailDescription>an IVOA Web Service shall provide a "availability" interface </failureDetailDescription>
    <testToDo>
      <InterfaceChecking>availability</InterfaceChecking>
    </testToDo>
  </testAssertion>

  <testAssertion id="R0210" type="required" enabled="true">
    <context>Support Interfaces</context>
    <assertionDescription>availability interface return testing</assertionDescription>
    <failureMessage>availability does not return an XML file conform to the VOSIAvailability schema</failureMessage>
    <failureDetailDescription>the "availability" interface shall return an XML document conform to the VOSIAvailability schema </failureD
    <testToDo>
      <OutputChecking type="XSD" interface="availability">link to VOSIAvailability schema </OutputChecking>
    </testToDo>
  </testAssertion>
</vowsbasicprofile>
```

# 5. Conformance claim

A conformant service shall be claimed if the service is compliant both with the WS-I Basic
Profile 1.1 and the WS-I Simple SOAP Binding 1.0 and VOSI 1.0. The wording of the
claim shall be:

**This web service is compliant with the IVOA Web Services Basic Profile 1.0 (or
later)**

This shall be understood to mean that at least all the mandatory rules are true. The list of
all the optional rules which are also verified should be also provided.

# 6. Changes from previous versions

- Take into account the last Support Interfaces document version
- Remove rules concerning non-mandatory interfaces
- Added relation to IVOA Architecture
- Reorganized document